\documentclass[BG]{copernicus}

\usepackage{graphicx}
\usepackage{tabulary}
\usepackage{epsfig}
\usepackage{amssymb}
\usepackage{multirow}
\usepackage{appendix}
\usepackage{natbib}
\newcommand{\beq}[1]{\begin{equation}\label{#1}}
\newcommand{\eeq}{\end{equation}}
\newcommand{\sub}[1]{_{\rm #1}}

\begin{document}

%\linenumbers

\title{The Habitable Zone of Inhabited Planets}

\author[1]{Jorge I. Zuluaga Callejas}
\author[2]{Juan F. Salazar }
\author[1]{Pablo Cuartas-Restrepo}
\author[3]{German Poveda}

\affil[1]{FACom - Instituto de F\'isica - FCEN, Universidad de Antioquia, Calle 70 No. 52-21, Medell\'in, Colombia}
\affil[2]{Escuela Ambiental, grupo GIGA, Facultad de Ingenier\'ia Universidad de Antioquia, Calle 70 No. 52-21, Medell\'in, Colombia}
\affil[3]{Escuela de Geociencias y Medio Ambiente, Universidad Nacional de Colombia, Medell\'in, Colombia}

\runningtitle{The InHZ}

\runningauthor{Zuluaga, Salazar, Cuartas, Poveda}

\correspondence{Jorge I. Zuluaga \\ (jzuluaga@fisica.udea.edu.co)}

\received{}
\pubdiscuss{} 
\revised{}
\accepted{}
\published{}

%% These dates will be inserted by the Publication Production Office during the typesetting process.

\firstpage{1}

\maketitle  

\begin{abstract}
 In this paper we discuss and illustrate the hypothesis that life
 substantially alters the state of a planetary environment and
 therefore, modifies the limits of the HZ as estimated for an
 uninhabited planet.  This hypothesis lead to the introduction of the
 Habitable Zone for Inhabited planets (hereafter InHZ), defined here
 as the region where the complex interaction between life and its
 abiotic environment is able to produce plausible equilibrium states
 with the necessary physical conditions for the existence and
 persistence of life itself.  We support our hypothesis of an InHZ
 with three theoretical arguments, multiple evidences coming from
 observations of the Earth system, several conceptual experiments and
 illustrative numerical simulations.  Conceptually the diference
 between the InHZ and the Abiotic HZ (AHZ) depends on unique and
 robust properties of life as an emergent physical phenomenon and not
 necesarily on the particular life forms bearing in the planet.  Our
 aim here is to provide conceptual basis for the development of InHZ
 models incorporating consistently life-environment interactions.
 Although previous authors have explored the effects of life on
 habitability there is a gap in research developing the reasons why
 life should be systematically included at determining the HZ limits.
 We do not provide here definitive limits to the InHZ but we show
 through simple numerical models (as a parable of an inhabited planet)
 how the limits of the AHZ could be modified by including plausible
 interactions between biota and its environment.  These examples aim
 also at posing the question that if limits of the HZ could be
 modified by the presence of life in those simple dynamical systems
 how will those limits change if life is included in established
 models of the AHZ.
\end{abstract}

\keywords{Habitable Zone; Habitability; Planetary Habitability and
  Biosignatures; Planetary Environments}

%%%%%%%%%%%%%%%%%%%%%%%%%%%%%%%%%%%%%%%%%%%%%%%%%%%%%%%%%%%%%%%%%%%%%%%%%
\introduction
\label{sec:Introduction}
%%%%%%%%%%%%%%%%%%%%%%%%%%%%%%%%%%%%%%%%%%%%%%%%%%%%%%%%%%%%%%%%%%%%%%%%%

\begin{flushright}
{\it ``It can scarcely be denied that the supreme goal of all theory
  is to make the irreducible basic elements as simple and as few as
  possible without having to surrender the adequate representation of
  a single' datum of experience''\\ \textbf{Albert Einstein, 1934}}
\end{flushright}
%FINAL 

The search for life outside the Solar System has nowadays reaching the
level where almost two thousand of extrasolar planets\footnote{For
  updates please refers to: http://exoplanet.eu/} and even more
exoplanet candidates \citep{Batalha13} have been discovered.  Assesing
the potential of these worlds to host surface liquid water (generally
assumed as the most important physical prerequisite for life) is
important and for that purpose it has been introduced the concept of a
``habitable zone'' (HZ) \citep{Dole64,Hart79,Kasting93}.
Traditionally, the definition of habitability has been mainly related
to planetary insolation, i.e. the equilibrium between the amount of
radiation a planet receives from its parent star and the energy the
planet radiates to space from its surface and atmosphere.  Planetary
insolation is supposed to determine the capacity of a planetary
environment to harbor surface liquid water and hence an evolving and,
probably more importantly, detectable biosphere.  A purely isolation
condition leads straightforwardly to the concept of a {\it Radiative
  Habitable Zone} (RHZ) \citep{Kasting93}, defined as the spherical
shell around a star where insolation, provided the planet have a dense
enough atmosphere, is compatible with surface liquid water and
probably with life as we know it.
%FINAL

Nowadays the definition of the HZ has trascended the pragmatical goal
of simply selecting which candidates in a exoplanetary survey could be
further studied.  Habitability has morphed in a complex and probably
more fundamental subject involving the unique properties of life and
its interaction with a dynamical planetary environment.
%FINAL

The definition the HZ, either obeying insolation or other complex
physical factors, assumes habitability as a necessary, although not
sufficient \textit{abiotic} condition for life.  However, and as the
observation of the Earth System (ES) suggests, habitability is not
only an abiotic prerequisite but an emergent property of a very
complex system involving the interaction among astrophysical,
geophysical and not less important biological factors \citep{Sagan72,
  Lovelock74, Margulis74, Walker81, Franck00a, Franck00b, Franck01a,
  Franck01b, Lovelock09, Rosing10}.
%FINAL

The effect that life has at determining the equilibrium state of a
habitable planet has been much less studied when compared with the
effects of many other abiotic factors.  This is especially true when
dealing with the estimation the limits of the HZ in extrasolar
planetary systems.
%FINAL

The key role of life in the environment has been widely discussed in
the literature of the ES
\citep{CaldeiraKasting92,Lenton02,Kleidon09a,Kleidon10,Kleidon12}.  In
the astronomical community, several works, including the seminal paper
by \citet{Kasting93}, have also posed and discussed the importance of
life at affecting planetary habitability (for a recent review
\citet{Kasting10} and references therein).  However, and as far as we
know it, the most consistent efforts attempting to include the effect
of biota on the long-term evolution of Earth-analogues' habitability
were those made more than a decade ago by S. Franck and collaborators
\citep{Franck99, Franck00a, Franck00b, Franck01a, Franck01b}. More
recently \citet{Dyke11,Honing13} has also explored the effects of life
in the evolution of several geophysical factors affecting planetary
habitability.  Despite these important efforts, a conceptual basis for
the general definition of a Habitable Zone for actually Inhabited
Planets, is still lacking.  This is clearly evidenced in the absence
of biotic factors in most, if not all recent habitability models.
%FINAL

In the case of the Abiotic HZ (AHZ), the lack of a statistically
significant number of observations able to confirm the HZ limits
predictions, required in the past the development of a solid
conceptual basis to support scientifically the theoretical models on
which predictions rely. Analogously the discussion of the role of life
on habitability requires the development of a general conceptual basis
before implementing specific models attempting to redefine the HZ
limits.
%FINAL

In order to develop that conceptual basis, we define and discuss here
the concept of a {\it Habitable Zone of Inhabited Planets} (hereafter
InHZ).  We support our definition in theoretical arguments based on
the understanding of the biota-environment interaction as observed in
the ES.  Since our Planet is the only habitable planet we know so far,
its properties are the only point of reference we have for this
construction.  This is analogous to the way as the Solar System rocky
planets (Venus and Mars) are used in the definition of the RHZ limits.
Furthermore, we show, through a conceptual experiments and numerical
simulations, how the InHZ limits would change with respect to the AHZ
in hypothetical inhabited planetary environments.
%FINAL

It is important to stress that our approach does not intend to give
(yet) numerical predictions about the extension of the InHZ in actual
planetary systems. Our aim here is to provide a general conceptual
basis for the development of models able to estimate these limits.
Moreover, by using numerical simulations of idealized inhabited
planets we just aim at posing the question that if limits of the HZ
could be modified by the presence of life on those simple dynamical
systems how will those limits change if life were included also in
more sophisticated models of the AHZ.
%FINAL

This paper is organized as follows: in Section \ref{sec:DefiningInHZ}
we define the InHZ and discuss it in the context of the well known AHZ
concept.  Section \ref{sec:TheoreticalArguments} is devoted to develop
the theoretical arguments that support the introduction of the InHZ.
In Section \ref{sec:TowardsQuantitativeModelInHZ} we present the
results of conceptual and numerical experiments of the
biota-environment interaction that illustrate quantitativeley the InHZ
definition.  Section \ref{sec:Discussion} is devoted to discuss the
limitations, open questions and consequences of pursuing the more
general InHZ as opossed to a the AHZ.  Finally in Section
\ref{sec:Conclusions} we summarize our proposal and draw some
conclusions and future prospects of this work.
%FINAL

%%%%%%%%%%%%%%%%%%%%%%%%%%%%%%%%%%%%%%%%%%%%%%%%%%%%%%%%%%%%%%%%%%%%%%%%%
\section{Defining the Habitable Zone of Inhabited Planets}
\label{sec:DefiningInHZ}
%%%%%%%%%%%%%%%%%%%%%%%%%%%%%%%%%%%%%%%%%%%%%%%%%%%%%%%%%%%%%%%%%%%%%%%%%

We define the \textit{Habitable Zone of Inhabited Planets} (InHZ) as
{\it the region (in space and time) where the complex interaction
  between life and its abiotic planetary environment is able to
  produce plausible equilibrium states with the necessary physical
  conditions for the existence and persistence of life itself}.  This
definition does not intend to replace the definition of the AHZ but to
extend it.
%FINAL

The reason why life is so important at determining the habitability of
a planet lies on its capacity to substantially alter its abiotic
environment.  For instance, on Earth, biogenic mass fluxes strongly
alter the atmospheric structure and composition at different spatial
and temporal scales \citep[e.g.][]{Beerling05,Poschl10}.  In the
simulations published by S. Franck and collaborators, almost 14 years
ago, the inclusion of some biotic feedbacks in the carbonate-sillicate
cycle modified substantially the life span of the biosphere in Earth
analogues.  These examples clearly suggest that life allows the
emergence of planetary equilibrium states that would not be
predictable if neglecting its effects.  In other words, a habitable
planet without life and the same planet actually inhabited by a
widespread biota are very different, especially in their potential to
give rise to plausible habitable equilibrium states.
%FINAL

An inhabited planet is a complex system comprising biotic and abiotic
components.  Taking away life and its powerful feedbacks is as
unnatural as removing liquid water or any other major component of the
system.  Removing key components of a complex system not only
perturbates the properties of the system but it could potentially
drive the system to qualitatively distinct equilibrium states.
%FINAL

It is worth noticing that our definition of an InHZ does not exclude
the requirement of other abiotic prerequisites.  To allow the
emergence of complex interactions between life and its abiotic
environment, a planet could also require a dense enough atmosphere,
complex geophysical processes (e.g. plate tectonics, volcanism) or a
protective magnetic field (see e.g. \citealt{Zuluaga13}).  Even in
this case, it has been recently recognized that life on Earth not only
has altered the evolution of the atmosphere and oceans.  Life could
also affect interior geological processes and other global planetary
factors \citep{Dyke11,Honing13}.  Together, all these evidences point
out to identify life not only as an important component of an
inhabited planetary environment but as a major geological force at all
levels.  This significant fact was already anticipated, at least in
the case of the ES, by Vladimir Vernadsky circa 1920.
%FINAL

Although at first sight the origin of life could be a problem for the
definition of an InHZ, this sort of ``egg-and-chicken paradox'' is
almost inevitable when dealing with complex systems.  Here, however,
it is interesting to notice that at defining the InHZ we would not
require to explain the appearance of the first forms of life in the
same way as the definition of the AHZ would not require to explain,
for instance, the appearance of the first drop of water.  We just need
to recall that the definition of the RHZ requires liquid water as a
prerrequisite for maintaining the Carbon-Sillicate cycle via
weathering processes \citep{Kasting93}.  Therefore, in the same way as
explaining the origin of liquid water is not mandatory to define the
RHZ, explaining the origin of life is not necesarily required to
define the InHZ.
%FINAL

%%%%%%%%%%%%%%%%%%%%%%%%%%%%%%%%%%%%%%%%%%%%%%%%%%%%%%%%%%%%%%%%%%%%%%%%%
\section{Theoretical arguments}
\label{sec:TheoreticalArguments}
%%%%%%%%%%%%%%%%%%%%%%%%%%%%%%%%%%%%%%%%%%%%%%%%%%%%%%%%%%%%%%%%%%%%%%%%%

\begin{flushright}
{\it ``We are only now beginning to acquire reliable material for
  welding together the sum total of all that is known into a whole
  [...] Some of us should venture to embark on a synthesis of facts
  and theories, albeit with second-hand and incomplete knowledge of
  some of them - and at the risk of making fools of
  ourselves''\\ \textbf{Erwin Schrodinger in ``What is Life?''
    (1992)}}
\end{flushright}

There are three key theoretical arguments supporting the idea that
habitability should not be assessed without including the influence of
life:
%FINAL

\begin{enumerate}

\item[(1)] Biota-environment feedbacks are likely to substantially
  alter the equilibrium states of any inhabited planet.
%FINAL

\item[(2)] The equilibrium state of a complex system cannot be
  predicted while neglecting one of its major components (in this case
  life).
%FINAL

\item[(3)] Living phenomena have unique properties, hardly mimicked by
  abiotic mechanisms and able to explain the maintainance of
  physically unstable states of inhabited planets.
%FINAL

\end{enumerate}

In the following paragraphs we develop in detail each argument and
present the observational and theoretical evidences supporting them.
%FINAL

%============================================================
\subsection{Argument 1: the power of biota-environment feedbacks} 
\label{subsec:ArgumentBiotaFeedbacks}
%============================================================

Life alters its environment and the environment constrains life. This
well-known two-way relationship implies the existence of
biota-environment feedbacks which can produce global scale effects as
life forms grow and reproduce \citep{Lenton03, Foley03}. These global
scale feedbacks will be an universal feature of planets inhabited by a
widespread biota.
%FINAL

Biota-environment feedbacks can strongly alter the physical conditions
that are regularly taking into account when defining abiotically the
HZ.  Thus, for instance the water and carbon content of the atmosphere
or the presence of clouds in the Earth, would not be the same if our
planet were uninhabited \citep{Lenton98, Lovelock95}.  We argue here
that in any inhabited planet the power of such biota-environment
feedbacks is too large to be neglected when definining the HZ
%FINAL

Life is based on biochemical reactions that continuously convert
inorganic substances stored in the environment into organic ones and
back.  Therefore, large biochemical fluxes of synthesis and
decomposition of organic substances are expected.  In the Earth the
power of these fluxes is such large that if they were not tightly
compensated, the environment could change dramatically in time-scales
of several tens of years \citep{Gorshkov04}.  Those changes could
bring the environment to a state incompatible with the existence of
life itself (Makarieva and Gorshkov, personal communication 2013)
%FINAL

One of the most noticeable biota-environment feedbacks regarding
habitability may be those related to clouds.  Water or carbon dioxide
clouds are key at determining the extension of the HZ (see
e.g. \citealt{Mischna00}, \citealt{Kitzmann10}). On Earth, water
clouds are a key component of the climate system, and its influence on
the equilibrium state of the environment is presumably large
\citep[see, e.g.][]{Ramanathan89}.  Land vegetation and phytoplankton
play an important role at controlling the amount of cloud condensation
nuclei (CCN) in the atmosphere \citep{Meskhidze06, Poschl10}, thereby
affecting the formation of clouds. The presence of native vegetation
may enhance the formation of clouds especially over certain areas of
the planet \citep{Lyons02}.  Even airborne microorganims living in the
middle-upper troposphere can work as biotic cloud condensation nuclei
\citep{Deleon13}.  In large scale natural forests such as the Amazon,
physical connections between clouds, rainfall and vegetation have been
also identified \citep{Andreae04, Bonan08}.
%FINAL

Biota-environment interactions on Earth are not restricted to the
effects on clouds.  Terrestrial biota also plays an important role
both in the hydrologic cycle \citep{Hutjes98} and in the global carbon
cycle \citep{Schimel95}.  It has been recently shown, for instance,
that terrestrial water fluxes are dominated by biological processes
(transpiration) rather than physical ones (evaporation)
\citep{Jasechko13}.  Moreover, a non negligible number of feedbacks
through which forests exerts strong effects on climate regulation have
been also identified \citep{Bonan08}.  For instance, it has even been
proposed that forest vegetation can interact with its surrounding
environment in ways that enhance conditions favorable for its own
existence \citep{Runyan12}. Natural forests may be responsible for a
biotic pump of atmospheric moisture driving the hydrologic cycle on
land \citep{Makarieva07,Makarieva10,Poveda14}.
%FINAL

Concerning the carbon-cycle, plant evolution on Earth, for example,
has strongly influenced the amount of CO$_2$ in the atmosphere at
geological timescales \citep{Beerling05}.  Other biological processes
such as the ecological success of calcareous plankton have driven
important changes in the global carbonate cycle \citep{Ridgwell05}.
These changes have had important implications on atmospheric and ocean
chemistry, and hence on the regulation and evolution of the ES at
geological timescales \citep{Ridgwell05}.  These evidences have lead
to recognize terrestrial biota as a key regulator of the atmospheric
chemistry and global Earth climate \citep{Arneth10}.
%FINAL

In summary biologically driven processes can significatively alter the
global biogeochemical and biogeophysical cycles and therefore, the
equilibrium habitable state of our planet would not be the same
without the effects of biota. Moreover, these effects exist as a part
of the complex Earth system, irrespectively of accepting that life
plays a determinant role at regulating the environment.
%FINAL

%============================================================
\subsection{Argument 2: Equilibrium States of Inhabited Habitable Planets}
\label{subsec:EquilibriumInhabitedHabitable}
%============================================================

The Earth functions as a whole complex system having physical,
chemical and biological coupled components.  It is not possible to
understand the functioning of the ES without considering it as whole
(\citealt{Houghton01}, p.784; \citealt{Rial04}).  Interestingly
\cite{Schellnhuber99} refers to modern scientific advances striving to
understand the ES as a whole and the development of new concepts on
this basis, as a ``second Copernican revolution''.  Turning from an
AHZ to an InHZ brings this ``revolution'' to the search for habitable
planets.
%FINAL

Although the knowledge of whether and how life provides an
``establishing'' influence on the ES remains elusive, there is no
doubt that biota plays a crucial role in the complex behaviour of the
system \citep[][p.69-70]{Steffen04}.  Consequently the plausible
equilibrium states of a planetary environment cannot be predicted,
especially if inhabited by a widespread biota, without taking into
account the role of life.  If our goal is to search for life in the
universe, the determination of the limits of the HZ, which is
essentially based on studying the plausible equilibrium states of the
system, must consider the role of life.
%FINAL

%FFFFFFFFFFFFFFFFFFFFFFFFFFFFFFFFFFFFFFFFFFFFFFFFFFFFFFFFFFFFFFFFFFFFFFFFFFF
\begin{figure*}[htp]
  \centering
  \includegraphics[width=0.8
    \textwidth]{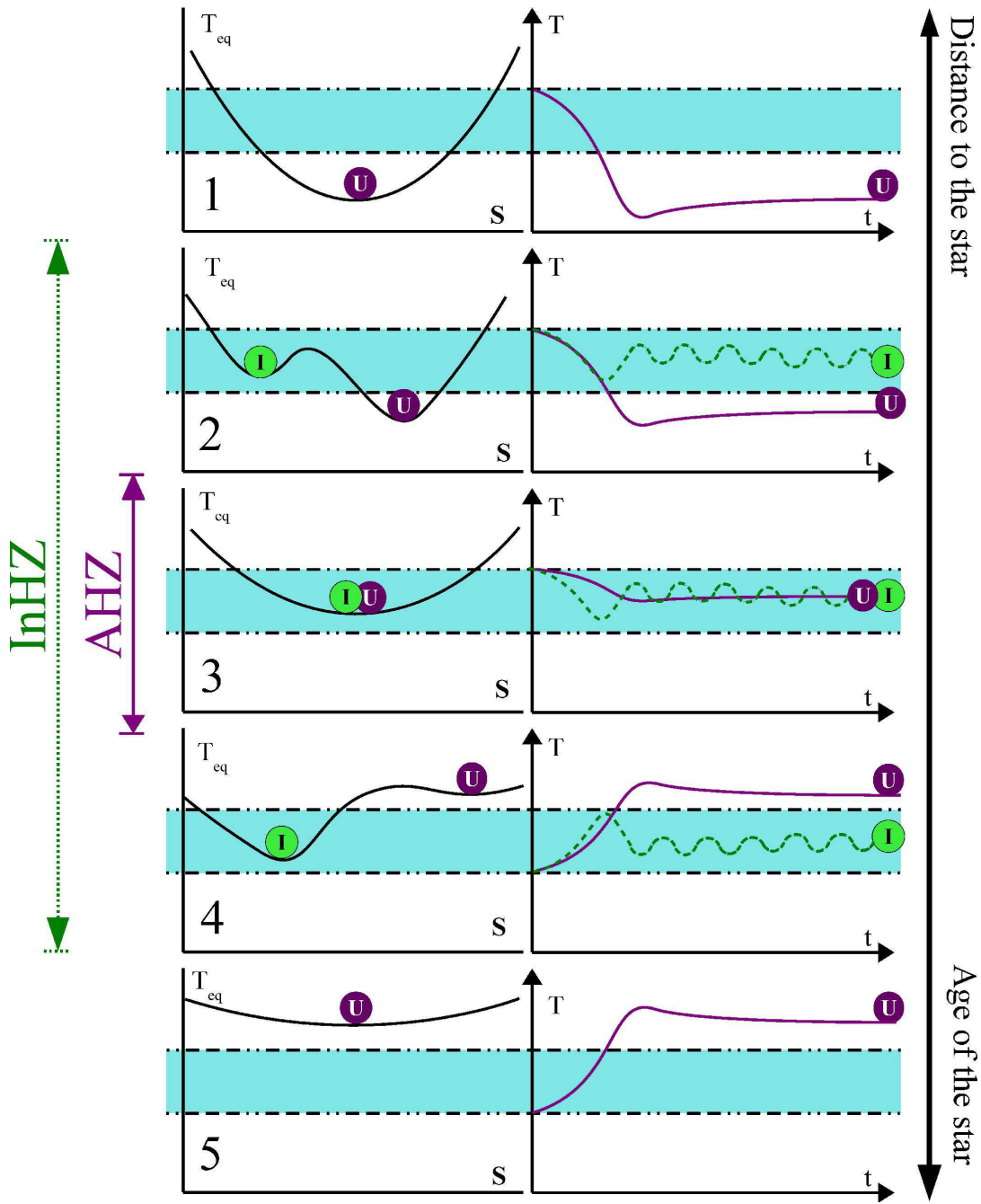}
  \caption{Conceptual representation of the environmental equilibrium
    states in uninhabited and inhabited planets. The potential wells
    represent stable atractors, and the ball the state of the system
    for an inhabited (I) or uninhabited (U) planet.  Right column
    shows the time evolution of global surface temperature until
    reaching equilibrium. Equilibrium states can be fixed points, or
    limit cycles characterized by temperature oscillations.  The range
    of temperature values within wich liquid water can exist is
    indicated by horizontal dashed lines. Stellar flux increases when
    moving from row 1 to 5 either in response to the evolution of the
    stellar luminosity or because we are closer to the
    star.}\label{fig:ConceptualInHZ}
\end{figure*}
%FFFFFFFFFFFFFFFFFFFFFFFFFFFFFFFFFFFFFFFFFFFFFFFFFFFFFFFFFFFFFFFFFFFFFFFFFFF

We illustrate the key differences between the AHZ and the InHZ in
Figure \ref{fig:ConceptualInHZ}.  We show there the equilibrium states
of uninhabited and inhabited planets.  The balls represents the states
of planetary environments, while the valleys or potential wells
depicted in the left column panels, represent stable atractors either
in the case of uninhabited (U) or inhabited (I) planets. An
uninhabited planet can be inside the HZ (i.e. the AHZ) if exist at
least one \textit{plausible} equilibrium state where the surface
temperature (and other environmental variables) is within the range of
values where liquid water can exist (shaded strip). This planet can be
also habitable if inhabited, although its equilibrium state can be
different (row 3).  Inhabited equilibrium states could be
characterized by biologically induced oscillations (a limit cycle)
rather than states characterized by almost constant values of the
environmental variables (fixed points) (in Section
\ref{subsubsec:LimitCycle} we provide a specific example of this
condition). Rows 1 and 5 illustrate those situations in which abiotic
conditions are prohibitively extreme for life, either because the
planet is too cold (row 1) or too hot (row 5). The important point
here is that there may exist intermediate conditions between cases 1,
3, and 5, which would be ``invisible'' without involving biotic
activity in the dynamic of the environment.  A coupled
biota-environment system may be able to find a habitable equilibrium
state in a planet which, if uninhabited, would be too cold (row 2) or
too hot (row 4) to be habitable. In this sense rows 2 and 4 would
represent planets within the InHZ.  It is important to stress that
these planets would be otherwise considered uninhabitable if the role
of life were ignored.  Since solar forcing increases from row 1 to row
5, these results can be interpreted as an stretching of the HZ towards
the InHZ.  This strectching occurrs either in time (going from row 1
to 5 solar luminosity increases) or in space (going from row 5 to 1
the distance to star increases).
%FINAL

In the examples depicted in rows 2 and 4 of Figure
\ref{fig:ConceptualInHZ} the presence of life in otherwise
uninhabitable planets allows the emergence of habitable conditions.
But there should also be another possibility.  Organisms producing a
de-stabilizing effect on the environment (disruptive organisms) could
also evolve and induce catastrophic events able to make permanently
uninhabitable a planet.  They could for instance change the
composition of the atmosphere inducing a runaway greenhouse effect or
bringing out radioactive elements from planetary interior sterilizing
permanently the surface. However, none of these scenarios would be
different than other catastrophic abiotic events (an asteroid impact
or a nearby supernova).
%FINAL

It should be then important to stress that the InHZ, as well as the
AHZ, are not defined depending on random events, either physical or
biological, but on the existence of plausible habitable equilibrium
states in the system under regular (non-transient) conditions.  In
order to further clarify the independence of habitability on random
events we propose the following general definition:
%FINAL

\begin{quote}
{\it A planet is within the HZ (either the AHZ or the InHZ) if under
  certain set of regular (non-transient) internal and external
  conditions, there exists at least one equilibrium state compatible
  with the existence and persistence of life.}
\end{quote}
%FINAL

Accordingly, if those conditions making plausible the existence of
habitable equilibrium states involve life, then the HZ would be an
InHZ.
%FINAL

%============================================================
\subsection{Argument 3: The unique properties of life} 
\label{subsec:UniquePropertiesofLife}
%============================================================

The role of life in the determination of the equilibrium state of the
ES has been widely discussed and extensively developed by two
complementary theories: Gaia \citep{Margulis74, Lovelock79, Lenton02}
and Biotic Regulation of the Environment (hereafter BR)
\citep{Gorshkov95, Gorshkov00, Gorshkov04}\footnote{It is interesting
  to recall that the original motivation of Gaia theory was precisely
  the search for extraterrestrial life \citep{Hitchcock67,
    Lovelock79}.  Therefore, the idea that life is somehow involved in
  the determination of the habitability of a planet has been implicit
  in the literature since the appearance of the aforementioned works
  and could even be traced back to the introduction of the biosphere
  concept by V.I. Vernadsky in 1926.}.  Independently and from a
purely physicochemical perspective the connection between life and the
regulation of the Earth environment has also been explored in the
studies of A. Kleidon and collaborators on the non-equilibrium
thermodynamics of the ES\citep{Kleidon10,Kleidon10b,Kleidon12}.
%FINAL

These three independent theoretical frameworks agree that the presence
of life on Earth plays a major role at determining the Earth's
physically unstable equilibrium state\citep{Margulis74, Lovelock79,
  Gorshkov95, Gorshkov00, Lenton02, Gorshkov04, Kleidon12}.  Moreover
and according to, for instance, BR, life makes the resulting unstable
state resilient and biotically stable (A. Makarieva and V. Gorshkov,
personal communication, 2013)
%FINAL

An interesting implication of the influence of life on the equilibrium
state of the ES is that the lifespan of Earth's biosphere can be
extended \citep{Lenton01}. This result is in agreement with the
independent simulations performed by S. Franck and collaborators
\citep{Franck00b}[e.g.] and with the argument presented in Section
\ref{subsec:EquilibriumInhabitedHabitable}
%FINAL

Together, all these theories and models support the notion that the
influence of life cannot be excluded when assesing the habitability of
inhabited planets.  This is because the presence of life confers to
the complex system (the inhabited planet) properties that (1) exerts a
non nengligible influence on the system's equilibrium and (2) are
hardly replaced or mimicked by even very complex abiotic factors.
%FINAL

\citet{Lenton98} developed several fundamental arguments about why
biota can be distinguished of other abiotic factors. In Lenton's own
words \textit{``in contrast to a dead world, the introduction of
  organisms brings an inherent tendency to stabilize conditions that
  are inhabitable by life''}.  This assertion alone agrees with the
intrinsic difference between an InHZ and an AHZ as argued here.
According to Lenton, three intrinsic properties of life drive an
inhabited planetary environment towards a self-regulated (stationary)
habitable state \citep{Lenton98}:
%FINAL

\begin{enumerate}

\item[(1)] Organisms alter the environment by taking and excreting
  energy and waste products.  At doing so, life can produce novel
  biogeophysical and biogeochemical feedbacks (e.g. feedback on growth
  and feedback on selection, see below) competing with and possibly
  dominating over the otherwise existing abiotic physicochemical
  feedbacks.
  %FINAL

\item[(2)] Organisms grow and multiply, potentially exponentially,
  leading to global positive feedback on the environment (more
  individuals means also a larger capacity to grow).  Growth tends to
  amplify any already existing biological feedback.
  %FINAL

\item[(3)] For each environmental variable there is a level or a range
  of values whereby a giving organism grows at a maximum rate.  This
  property gives rise to the existence of positive and negative
  biota-environment feedbacks around the optimum values of the
  environmental variables.  With enough biological amplification, the
  interplay of those positive and negative feedbacks tends to
  stabilize the whole system.
  %FINAL

\end{enumerate}

Life is also unique because it can produce two kind of feedbacks not
present on abiotic systems, namely feedback on growth and feedback on
selection.  These novel feedbacks can have a large effect on the
regulatory capacity of the system \cite{Lenton98,Lenton04}.  The
feedback on growth occurs when an organism induces changes in the
environment that affects in the same way the growth of every competing
organism, so no selection force is induced.  The feedback on selection
occurs when the changes an organism introduces in the environment
affect distinctly each specie creating a selection force: organisms
which are affected in the sense that their growth is reduced under the
modified conditions tend to disappear. On the other hand, the species
being favored by the change are selected and stay alive.  Feedback on
growth and feedback on selection are hardly found in other abiotic
complex systems.
%FINAL

Independently, V.G. Gorshkov and A. Makarieva \citep{Gorshkov01,
  Gorshkov02, Gorshkov04, Makarieva06} identify other two unique
properties of living systems confering life unparallel capacities with
respect to the abiotic world \citep{Gorshkov04}:
%FINAL

\begin{enumerate}

\item[(4)] ``Living matter features a level of orderliness
  incomparably higher than than the surrounding environment''
  \citep{Gorshkov04}.
  %FINAL

\item[(5)] ``Life supports its orderliness in a way unprecedented in
  the inanimate world: by competitive interaction'' \citep{Gorshkov04}
  %FINAL

\end{enumerate}

According Gorshkov and Makarieva the large effect of life on the
regulation of the environment (biotic regulation), i.e. the
maintanance of the unstable habitable equilibrium state, results from
the correlated functioning of organisms that form local ecological
communities \citep{Gorshkov04}.  These correlations depend on
information stored in the genomes of biological species. In non-living
open physical systems a similar link between their regulatory capacity
and information can be established.  In this case information is
associated to the degree of orderliness in the system (number of
available degrees of freedom).  \citet{Gorshkov01} have estimated that
the amount of information stored in living systems is 24-25 orders of
magnitude larger than that of open physical systems observed in the
environment.  Such a huge difference stems from the fact that memory
in living cells (DNA and proteins) are microscopic, while memory and
self-organization units of abiotic processes are all macroscopic.  A
direct consequence of this difference is that regulatory capacity of
living system is much larger than that of non-living systems.
According to the second law of thermodynamics, to maintain a level of
orderliness (amount of information) comparable to living phenomena, an
abiotic system requires a prohibetively large amount of external
energy.  
%FINAL

Moreover, living systems can maintain such high level of orderliness
during, in principle, unlimited periods of time.  This is achieved
through competitive interaction, another unique feature of life.
Relatively disordered living systems (sick systems) are continually
replaced by highly ordered ones (healthy systems). Although the level
or orderliness of any individual organism inevitably decays following
the second law of thermodynamics, the global orderliness in a
population of organisms can be maintained since disordered organism
are continually replaced by highly ordered ones.
%FINAL

From a different perspective A. Kleidon, among other authors, have
highlighted two additional key properties of life regarding its
significant influence on the non-equilibrium thermodynamics of the ES:
%FINAL

\begin{enumerate}

\item[(6)] Life as any other complex systems with sufficient degrees
  of freedom, obeys the Maximum Entropy Principle, maintain a steady
  state at which entropy production is maximized
  \citep{Schrodinger92,Lorenz02,Ozawa03,Kleidon05}.
  %FINAL

\item[(7)] Life (in particular photosynthetic living organisms)
  ``generates substantial amounts of chemical free energy which
  essentially skips the limitations and ineffeciencies associated with
  the transfer of power within the system'' \citep{Kleidon10}.
  %FINAL

\end{enumerate}

After analysing the entropy balance of the Earth as a coupled,
hierarchical and a non-equilibrium thermodynamic system it has became
apparent that a widespread biota plays a driving role at generating
and maintaining the habitable equilibrium state of the system
\citep{Kleidon10,Kleidon10b,Kleidon12}.  There is no reason to think
that this driving role will not be also present on other inhabited
planets.  If so, this theoretical result agrees with the notion that
the equilibrium state of an inhabited planetary environment cannot be
predicted without taking into account the biotic activity.
%FINAL

It is worth noticing here that properties (1)-(7), although identified
after studying Earth's life, are not tightly coupled to a specific
model of life.  Instead, they are rooted on very basic physical
principles valid elsewhere in the Universe. In other words, the
definition of an InHZ is supported on general properties of life as a
physical complex phenomenon and not only on life as we know it on
Earth.
%FINAL

%%%%%%%%%%%%%%%%%%%%%%%%%%%%%%%%%%%%%%%%%%%%%%%%%%%%%%%%%%%%%%%%%%%%%%%%%
\section{Towards a quantitative model of the InHZ}
\label{sec:TowardsQuantitativeModelInHZ}
%%%%%%%%%%%%%%%%%%%%%%%%%%%%%%%%%%%%%%%%%%%%%%%%%%%%%%%%%%%%%%%%%%%%%%%%%

The conceptual basis constructed over the preceeding definitions,
theoretical arguments and observational evidence, is essentially aimed
supporting the construction of quantitative specific models of the HZ
for Inhabited Planets.  Although models including biota-environment
interactions have been extensively developed and applied to study the
ES (see e.g. \citet{Lenton01} and references there in) the case of
potentially inhabited extrasolar planets has been much less studied.
%FINAL

Here we present two examples of how can we asses the estimation of the
InHZ limits.  Limited by the huge complexity of fully-fledged models,
we present here two simple albeit illustrative examples.  A conceptual
experiment showing how an otherwise improbable environmental state,
becomes plausible under the complex dynamics of an inhabited
environment.  This conceptual experiment is aimed to illustrate and
reinforce the argument in section
\ref{subsec:EquilibriumInhabitedHabitable}.  Then we present
simulation results of an idealized inhabited environment, a recent
variant of the ``Daisyworld''.  There we also review the results from
many other variants of this model to the light of the InHZ definition.
%OK

%============================================================
\subsection{The Inhabited Greenhouse-Albedo Cycler}
\label{subsubsec:LimitCycle}
%============================================================

The dynamics of complex systems can exhibit equilibrium states
(attractors) that are fixed points or limit cycles.  We refer to limit
cycles as states characterized by non negligible oscillations.  Most
of the models of the AHZ (e.g. \citealt{Kasting93} or
\citealt{Selsis07}) rely on fixed point planetary equilibrium states.
This is the case, for instance, of habitable planets in the innermost
edge of the RHZ in \citet{Selsis07}.  Those planets remain habitable
due to a constant 100\% covering of water clouds.
%FINAL

But, what would happen if the habitable state of an inhabited planet
is a limit cycle instead of a fixed point?  The complex interaction
between life and its abiotic environment very oftenly gives rise to
natural oscillations in the environment.  As a matter of fact limit
cycles are the rule and not the excepton in life-bearing dynamical
systems (see e.g. \citealt{Nicolis73}). One good example of this is
the seasonal cycle of CO$_2$ in Earth's atmosphere (see e.g. the
``Keeling curve'' \citealt{Keeling08}) which is driven by seasonal
changes in Earth's vegetation \citep[see, e.g.][]{Keeling96}.
%FINAL

Could an inhabited planet under plausible oscillating conditions be
able to stretch out the AHZ limits?.  In the following paragraphs we
provide an example of such a plausible enhanced inhabited habitability
equilibrium state.
%FINAL

\begin{figure}[t]
%   \vspace*{2mm}
  \centering
  \includegraphics[height=9.3cm]{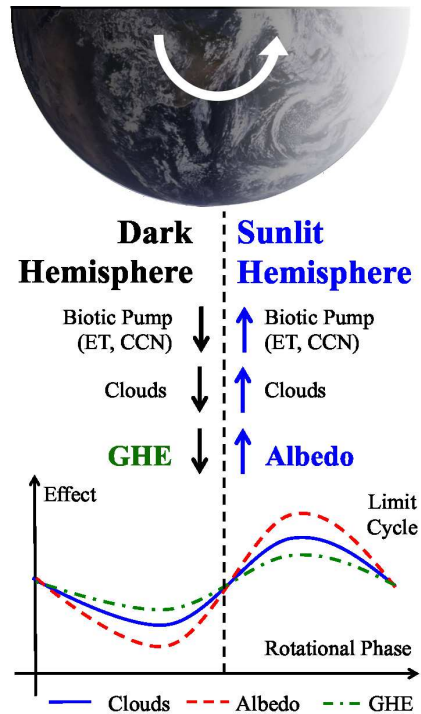}
  \caption{The ``Inhabited Greenhouse-Albedo Cycler'', a hypothetical
    inhabited planet maintained by the complex interaction between
    life (referred here as the biotic pump) and its environment in a
    limit cycle where clouds mostly cover only the sunlit hemisphere.
    As the planet rotates clouds and water vapor moves through the
    planetary surface creating enhanced habitability conditions.
    While the sunlit hemisphere is cooled by the higher albedo of
    clouds the dark hemisphere radiates more easily the accumulated
    heat to space due to a reduction in the Greenhouse effect (GHE).
    ET refers to Evapo-Transpiration and CCN to Cloud Condensation
    Nuclei, two of the by-products of biotic activity that strongly
    affects the formation of water clouds.}\label{fig:InHZLimitCycle}
\end{figure}

Let us assume a hypothetical planet that remains mostly covered by
water clouds only in the sunlit hemisphere while permanently uncovered
by them in the opposite dark hemisphere (see Figure
\ref{fig:InHZLimitCycle}).  If the planet is rotating at a faster rate
than the orbital angular velocity (i.e. a non-tidally-locked planet),
it will reach an oscillating equilibrium state, i.e. a limit cycle
(lowest diagram in Figure \ref{fig:InHZLimitCycle}).
%FINAL

%FFFFFFFFFFFFFFFFFFFFFFFFFFFFFFFFFFFFFFFFFFFFFFFFFFFFFFFFFFFFFFFFFFFFFFFFFFF
% \newpage
%FFFFFFFFFFFFFFFFFFFFFFFFFFFFFFFFFFFFFFFFFFFFFFFFFFFFFFFFFFFFFFFFFFFFFFFFFFF

In a dynamical state, clouds are continuously formed and destroyed in
the atmosphere.  This implies vertical transport of water.  Therefore,
the presence of clouds in a given place in the atmosphere implies also
the presence of water along the entire atmospheric column.  In such a
state, less (more) clouds will necessarily imply a weaker (stronger)
greenhouse effect.  In our hypothetical planet the absence of clouds
in the dark hemisphere makes it plausible to assume that the IR
opacity of the atmosphere on that side will be less than it would be
if fully covered by clouds.  Greenhouse effect (GHE) will then be
reduced and heat will more easily escape from the dark than from the
sunlit hemisphere. On the other hand, the permanently cloud covered
sunlit hemisphere, although more opaque to IR, will also have an
increased albedo.  If we assume that the cloud-forming effects of life
(evapotranspiration, ET, and the release of organic cloud condensation
nuclei, CCN) only affect the formation of low altitude clouds, a sky
permanently covered with this type of clouds will produce a net
cooling effect on the sunlit hemisphere.
%FINAL

At a given point in the planetary surface, the cloud covering and
water content and consequently the albedo and greenhouse effect, will
oscillate as the planet rotates (we assume the planet is not a
tidally-locked planet).  The net effect will be to maintain every
point as cool as possible combining the most favorable effect at the
right hour of the day: higher albedo at noon, and reduced greenhouse
effect at midnight.  As a results our planet would potentially reach
in this oscillating state, average surface temperatures lower than a
planet fully covered by clouds.  
%FINAL

In terms of HZ limits, at distances where even 100\% cloud covered
planets are uninhabitable, fast rotating inhabited planets with a half
cloud covered hemispheres controlled by the complex interaction
between life and the environment, could be cool enough to support
life.  In other words the HZ of this ``Inhabited Greenhouse-Albedo
Cyclers'' could be strecthed out towards the star.  Figure
\ref{fig:InHZLimitCycle} summarize schematically the dynamics of this
hypothetical inhabitabed habitable planet.
%FINAL

In summary, under the same external forcing where abiotic (fixed
point) equilibrium states are uninhabitable, an inhabited planet could
achieve a habitable state through a plausible limit cycle resulting
from the interaction between abiotic and biotic processes.
%FINAL

%============================================================
\subsection{A toy model of the InHZ}
\label{subsubsec:DaisyWorld}
%============================================================

One of the best known and extensively used toy model of an inhabited
planet is the Daysiworld (DW) model. Originally introduced by
\citet{Watson83}, the DW model is intended to simulate the dynamics of
a hypothetical simplified inhabited environment.  In the model,
instead of numerous variables describing the state of what otherwise
would be a very complex system, only one state variable is considered:
surface temperature.  Moreover the biota is greatly simplified to
contain only two types of living organisms: black and white daysies
\citep{Lenton01b}.
%FINAL

DW model was originally conceived as a parable with possible
implications for Earth science.  Its spirit is that of answering
``what if...?''  questions regarding the interactions between life and
the environment \citep{Lenton01b}.  We will use here DW models in the
same spirit.  Thus, for instance, since the results of our DW model
shows that the interaction between biota and its environment is able
to modify the limits of the AHZ what would be life able to do, in more
complex models.  We will come back on this question later on.
%FINAL

Many variants of the original DW have been developed.  Each of them
have been intended to study different aspects of the role of life at
regulating the environment \citep{Lenton01b, Wood08}.  Although
well-known and widely described in literature, we will briefly present
here the general features of the DW model, focusing especially on the
recent variant developed by \citet{Salazar09}.  In their variant
\citet{Salazar09} introduced the interaction between life and the
hydrological cycle.  As we will show here, this key interaction is the
determinant driver of planetary habitability.
%FINAL

%FFFFFFFFFFFFFFFFFFFFFFFFFFFFFFFFFFFFFFFFFFFFFFFFFFFFFFFFFFFFFFFFFFFFFFFFFFF
%FIGURE 2: 
\begin{figure}[t]
  \centering
  \vspace*{2mm}
  \includegraphics[width=8.3cm]{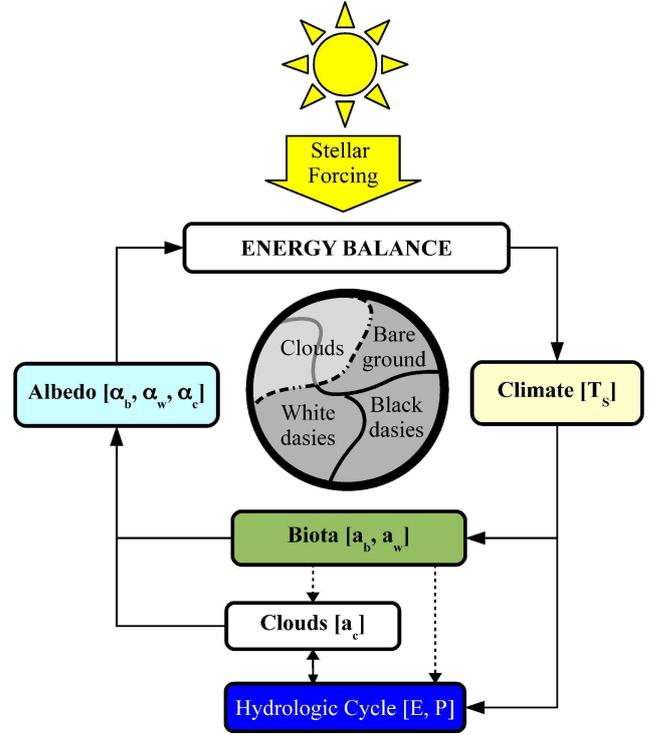}
  \caption{Schematic representation of a DW model including clouds and
    a hydrologic cycle \citep{Salazar09}. Components of the system
    (subsystems and processes) are indicated inside rectangles.  Basic
    properties associated to each component are written in brackets
    (see equations in the main text).  Feedbacks between components
    are indicated with arrows.  Dashed arrows going from biota to
    clouds and to hydrologic cycle are hypothetical feedbacks not
    included yet in the DW variant studied
    here.}\label{fig:SchematicDW}
\end{figure}
%FFFFFFFFFFFFFFFFFFFFFFFFFFFFFFFFFFFFFFFFFFFFFFFFFFFFFFFFFFFFFFFFFFFFFFFFFFF

The DW lies on the surface of a hypothetical Earth-sized planet,
orbiting a star that provides all the required energy for an inhabited
environment.  The surface of the planet is partially covered by two
different species of ``daisies'' differing, among other properties, in
albedo. The original DW and many of its variants considers only
``black'' and ``white'' daisies (dark and light vegetated types of
land cover).  In other variants additional species having intermediate
albedos are also introduced. At any time the surface of the planet is
covered by variable fractions of daisies and/or bare ground.
%FINAL

The most important environmental variable in the model, the surface
temperature $T_s$, is described by the global energy balance and the
population-dynamics of the simplified biota (daisies). In the variant
of \citet{Salazar09} variant, the role of clouds and that of a global
hydrologic cycle are also included in this balance.  On a planet
covered by fractions $a_c$, $a_w$, $a_b$ and $a_d$ of clouds, white
and black daisies and bare-ground, respectively, the energy balance is
granted if:

\beq{eq:energy_balance}
c_{p}\frac{dT_s}{dt}={{S S_{\odot}}\over 4}
(1-\alpha_{c}a_{c})(1-\alpha_{s})+\sigma T_{c}^{4} a_{c}-\sigma
T_s^{4}.
\eeq

Here $S$ is the stellar energy flux measured in units of the solar
flux at Earth distance (Solar constant, $S_{\odot}$), $c_{p}$ is the
planetary mean heat capacity or thermal inertia and $\sigma$ is the
Stefan-Boltzmann constant. $\alpha_{s}$ and $\alpha_{c}$ are the
albedo of planetary surface and clouds, respectively.  $T_{c}$ is the
temperature of clouds which is related to $T_s$ through the
atmospheric lapse rate \citep{Salazar09}.  Total surface albedo
depends on the area fractions of black daisies, white daisies and bare
ground, whose albedos are denoted by $\alpha_{b}$, $\alpha_{w}$, and
$\alpha_{d}$ respectively.  We are adopting here for the stellar
adimensional irradiance $S$, the notation commonly used in
habitability literature instead of that used in DW literature where
$L$ is used to denote this quantity.  It should be noticed that
removing clouds ($a_c=0$) and giving the planet a null thermal inertia
($c_p=0$) leads to an exact radiative balance between the net incident
short-wave and outgoing long-wave radiation, i.e. $(S S_{\odot}/ 4)
(1-\alpha_{s}) = \sigma T_s^{4}$ which correspond to the energy
balance equation in the original DW model \citep{Watson83} and most of
its variants.
%FINAL

The evolution of daisies populations is given by the growth equations
\citep{Carter81},
%FINAL

\beq{eq:daisies_areas}
    \frac{d a_{i}}{d t}=a_{i}[(1-a_{w}-a_{b})\beta_{i}-\gamma],
\eeq

where the subscripts, $i=b,w$ refer to black ($b$) and white ($w$)
daisies.  Here $\beta_{i}$ is the growth rate of the $i$-daisies which
is a function of the planetary temperature and other parameters
conceptually related to their biology (e.g. tolerance to high
temperatures, adaptation, symbiosis); $\gamma$ is the mortality rate
commonly assumed as constant for both type of daisies. Different
variants of DW use modified version of Eq. (\ref{eq:daisies_areas})
and different functions $\beta_i$ intended to simulate certain aspects
of biological dynamics such as biodiversity \citep{Lovelock92},
adaptation \citep{Lenton00, Robertson98}, different forms of
competition \citep{Cohen00}, or symbiosis \citep{Boyle11}.
%FINAL

The hydrologic cycle, included for completeness in our description of
the model and first introduced by \citet{Salazar09}, is described with
the mass balance equation:
%FINAL

\beq{eq:evap_precip}
\frac{da_{c}}{dt}=E-P,
\eeq

where $a_c$ is assumed proportional to the atmospheric water content
and $E=(1-a_{c})E_p$ and $P=a_{c}P_p$ are the actual
evapotranspiration and precipitation rates. For a detailed description
of these factors please refer to \cite{Salazar09}.
%FINAL

Figure \ref{fig:SchematicDW} depicts schematically the feedbacks at
play in a DW model including clouds and a hydrologic cycle.  The only
external forcing in the model is the stellar insolation.  In the inner
loop (energy balance - climate - biota - albedo) the energy balance
between incoming and outgoing radiation determines surface planetary
temperature which in turn influences daisies population (biota)
changing planetary albedo.  Albedo finally enters into the energy
balance closing the loop.  When including the effect of the hydrologic
cycle and clouds, surface temperature also affects the exchange of
water between the surface and the atmosphere through precipitation and
evapotranspiration.  The presence of clouds in the atmosphere depends
on such processes and viceversa.  Clouds influences also planetary
albedo (and greenhouse effect) which in turn determine the energy
balance at the top, closing the loop. Since habitability depends
directly on surface temperature, habitable equilibrium states in DW
models will depend on the complex relationship among all these
feedbacks.  For completeness, two hypothetical feedbacks (dashed
arrows in Figure \ref{fig:SchematicDW}), not yet included in any DW
variant, have been also depicted in our schematic representation.
These feedback arises from the biota-cloud and biota-hydrologic cycle
interactions, and they have been mentioned in detail in Section
\ref{subsec:ArgumentBiotaFeedbacks} in the case of the ES.  They are
recognized as potential key drivers of regulatory dynamics in a
inhabited planet and should be included in future improved
biota-environment models.
%FINAL

\begin{figure}[t]
  \vspace*{2mm}
  \centering 
  \includegraphics[width=8.3 cm]{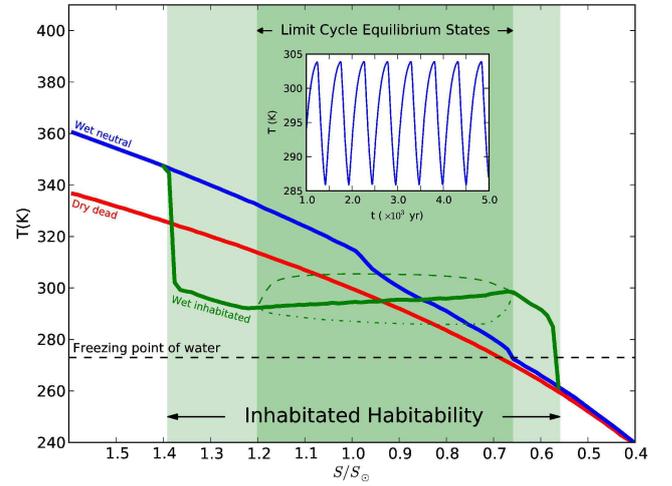}
  \caption{Equilibrium temperature versus stellar radiative forcing in
    a DW model including clouds and a hydrological cycle (green solid
    curve).  For comparison the equilibrium temperature of the planet
    when uninhabited (red and blue lines are also included) .  In this
    particular case the presence of life places the outer limit of
    InHZ further away from the star.  Compare the place where green,
    blue and red lines crosses the freezing temperature of water
    (black dashed line) at low stellar
    insolations.}\label{fig:ResultsDW}
\end{figure}

To illustrate the emerging properties of a DW with a hydrological
cycle we show in Figure \ref{fig:ResultsDW} a typical result of
solving the DW equations for different stellar insolations
\citep{Salazar09}.  We plot there the equilibrium temperature as a
function of the input stellar irradiance for the model also depicted
in Figure 7 of \citealt{Salazar09}.  In this case we have assumed a DW
covered by clouds with a 0.6 albedo located at 4 km above planetary
surface.  This clouds resembles the mid altitude clouds in the Earth.
We have verified that the conclussions drawn from this example will
not change too much when changing other model parameters.
%FINAL

%FFFFFFFFFFFFFFFFFFFFFFFFFFFFFFFFFFFFFFFFFFFFFFFFFFFFFFFFFFFFFFFFFFFFFFFFFFF
% \newpage
%FIGURE 3: 
%FFFFFFFFFFFFFFFFFFFFFFFFFFFFFFFFFFFFFFFFFFFFFFFFFFFFFFFFFFFFFFFFFFFFFFFFFFF

In the figure the difference between the equilibrium surface
temperature when dead, i.e. uninhabited (curves marked as wet neutral
and wet dry) and when inhabited, is notorious, especially in the solar
forcing range $S/S_\odot$=0.56-1.39.  This is precisely what we can
call here the Inhabited Habitable Zone of the system.  In this range
of solar forcing, surface temperatures stay regulated around the
temperatures where the growth of daisies is optimum ($T_s=20-25$ C).
%FINAL

Without life a stellar forcing of for example $S/S_\odot=0.6$ will
produce an equilibrium state (with and without clouds and greenhouse
effect) characterized by surface temperatures below the freezing point
of water (uninhabitable planet).  With the same stellar forcing the
biota-environment system reaches an equilibrium state with
temperatures almost 20 degrees above the freezing point of water.
%FINAL

More interestingly is the interval of $S/S_\odot$=0.66-1.20 the
equilibrium state of the system in that range is not a fixed point but
a limit cycle characterized by temperature oscillations with a
significant constant amplitude and mean value (inset
panel). Temperatures in these oscillatory states are represented in
the figure by their mean (wet inhabited solid line), maximum (dashed
line) and minimum (dash-dotted line) values.  The inset figure shows
the oscillations corresponding to $S/S_\odot=1$.  Since the occurence
of such oscillations depends on the biota-environment interactions,
the range of values of $S$ where oscillations arise, is inside the
InHZ.  It is interesting to notice however that not all the
equilibrium states inside the InHZ are limit cycles.  For example at
minimum and maximum values of $S$ the equilibrium states are actually
fixed points.  This behavior arises from the fact that at those
extremes the interspecific competition dissapears.  This is due to the
fact that only one the species is inhabitating the planet.  Table 2 in
\cite{Salazar09} shows that for a wide range of model parameters the
equilibrium states are limit cycles.  This confirms the idea that such
type of oscillating behavior in system with complex biota-environment
interactions are the rule rather than the exception. The noticeable
prevalence of limit cycles in DW is in agreement with the previous
results by \cite{Nevison99} and support the hypothesis behind the
conceptual experiment discussed in Section \ref{subsubsec:LimitCycle}.
%FINAL

\begin{figure}[t]
  \vspace*{2mm}
  \centering
  \includegraphics[width=8.3cm]{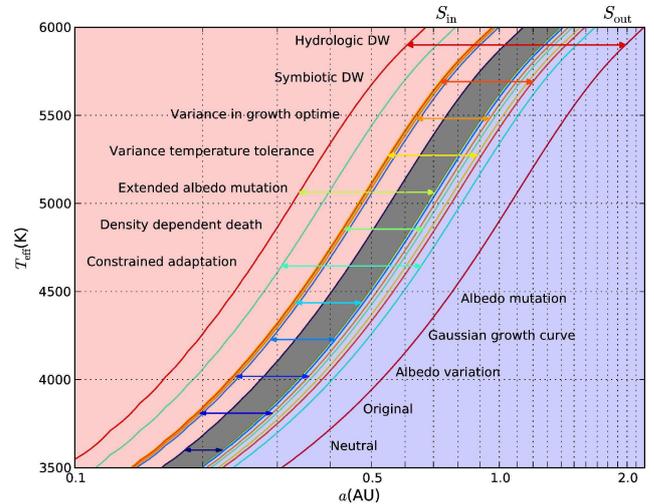}
  \caption{Inhabited Habitable Zones for different variants of the DW
    model (see Table \ref{tab:SummaryResultsDW}).  Curves indicate the
    distance $a$ to a main sequence star with effective temperature
    $T\sub{eff}$ where equal incoming flux $S/S\sub{\odot}$ is
    received in the inner and outer InHZ edges.  Labels summarize the
    criteria distinguishing each variant. The length of the double
    arrow in the middle of each strip represent the span $S\sub{R}$ of
    the InHZ.}\label{fig:DWInHZ}
\end{figure}

To summarize the evidence coming from a wide diversity of DW models
supporting the InHZ concept, we present in Table
\ref{tab:SummaryResultsDW} the limits of the HZ as calculated using
different assumptions about biota properties and its interspecific
interactions.  This table is a modified extension of that published by
\cite{Lenton01b}.  A graphical representation of these results are
presented in Figure \ref{fig:DWInHZ}.  We have used there the common
graphical representation of the circumstellar habitable zone
representing curves of constant $S$ in a plane of $T\sub{eff}$ vs. $a$
(stellar effective temperature vs. planet-star distance).  It is
important to stress that although using the same graphical
representation, a direct comparison among Figure \ref{fig:DWInHZ} and
a similar one representing the limits of the AHZ as calculated for
example with 1-D atmospheric models (e.g. \cite{Kopparapu13}) is not
straightforward.  It is worth to recall here DW models use very
simplified (if not unexistent) models for the atmopsheric response to
the incoming stellar radiation and depend on even simpler models of
what would be very complex biota-environment interactions.  The
Inhabited Habitable Zones in Figure \ref{fig:DWInHZ} are a parable of
real InHZs in the same way as DW is a parable of the ES.
%FINAL

The results compiled in Table \ref{tab:SummaryResultsDW} and
represented in Figure \ref{fig:DWInHZ} clearly show that the limits of
the HZ are sensitive first to the presence of life (the width of the
HZ in planets with biota-environment interactions are at the least two
times wider than that corresponding to a neutral planet), and second
to the particular properties of life and its interaction with the
environment.  Thus, for example giving some adaptation capabilities to
daisies (see {\it Constrained Adaptation} InHZ) widens the HZ span
with respect to other DW models where the intrinsic properties of life
are constant and independent of the environment (see e.g. {\it
  Original} InHZ).  These results lend support to the idea that the
InHZ could be significantly different (and probably much wider) than
the AHZ.
%FINAL

%TTTTTTTTTTTTTTTTTTTTTTTTTTTTTTTTTTTTTTTTTTTTTTTTTTTTTTTTTTTTTTTTTTTTTTTTTTTTTTTT
% \newpage
\begin{table*}[t]
% \scriptsize
\centering
\caption{Inner limit $S\sub{in}$, outer $S\sub{out}$ limit and span
  $S\sub{R}=S\sub{out}-S\sub{in}$ of the InHZ for different variants
  of the DW model. Adapted from
  \cite{Lenton01b}.}\label{tab:SummaryResultsDW}
\vspace{5 mm}
\begin{tabular}{llccc}  
    \hline 
    \textbf{Reference} & \textbf{Criterion} &  $S_{out}$ & $S_{in}$ & $S_R$  \\ 
    \hline\hline
    \citealt{Watson83} & Neutral Daisy &  $0,74$ & $1,11$ &  $0,37$ \\
		       & Original Daisyworld &  $0,68$ & $1,50$ & $0,82$  \\
    \hline
    \citealt{Lovelock89} & Albedo variation  &  $0,68$ & $1,50$ & $0,82$ \\
    \hline
    \citealt{Harding96} & Gaussian growth curve  &  $0,72$ & $1,45$ & $0,73$ \\
    \hline
    \citealt{Lenton98} & Albedo mutation &  $0,74$ & $1,50$ & $0,76$ \\
    \hline
    \citealt{Lenton00} & Constrained adaptation  &  $0,51$ & $2,32$ & $1,81$ \\
    \hline
    \citealt{Lenton01b} & Density dependent death  &  $0,65$ & $1,53$ & $0,88$ \\
		        & Extended albedo mutation  &  $0,74$ & $3,20$ & $2,46$ \\
			& Variance in Temperature tolerance  &  $0,60$ & $1,57$ & $0,97$ \\
			& Variance in growth optima  &  $0,68$ & $1,50$ &  $0,82$ \\
    \hline
    \citealt{Boyle11} &  Symbiotic DW increasing luminosity &  $0,57$ & $1,55$ & $0,98$ \\
    \hline 
    \citealt{Salazar09} & Hydrologic Daisyworld  & $0,30$ & $3,26$ & $2,96$ \\
    \hline\hline
\end{tabular}
\end{table*}
%TTTTTTTTTTTTTTTTTTTTTTTTTTTTTTTTTTTTTTTTTTTTTTTTTTTTTTTTTTTTTTTTTTTTTTTTTTTTTTTT

%FFFFFFFFFFFFFFFFFFFFFFFFFFFFFFFFFFFFFFFFFFFFFFFFFFFFFFFFFFFFFFFFFFFFFFFFFFF
% \newpage
%FIGURE 4: 
%FFFFFFFFFFFFFFFFFFFFFFFFFFFFFFFFFFFFFFFFFFFFFFFFFFFFFFFFFFFFFFFFFFFFFFFFFFF

%%%%%%%%%%%%%%%%%%%%%%%%%%%%%%%%%%%%%%%%%%%%%%%%%%%%%%%%%%%%%%%%%%%%%%%%%
\section{Discussion}
\label{sec:Discussion}
%%%%%%%%%%%%%%%%%%%%%%%%%%%%%%%%%%%%%%%%%%%%%%%%%%%%%%%%%%%%%%%%%%%%%%%%%

The theoretical arguments, conceptual experiments and numerical models
presented here have esatblished a minimum conceptual framework on
which justifying and building planetary habitability models including
the unavoidable effect of life itself.  However, an in depth approach
to the estimation of a realistic InHZ, will require the assesment of
several key issues not discussed yet.  Although an exhaustive
enumeration and discussion of the the many aspects involved in this
problem is certainly out of the scope of this paper, we will try to
summarize here some of the most important open issues that should be
addressed in forecoming papers.
%FINAL

At which extent the limits of the InHZ will depend on the very
specific properties of life inhabiting a given planetary environment?
In other words, are the InHZ limits different for each type of
organisms?  Will each form of life define a different InHZ even within
the same planetary system?
%FINAL

As discussed in Section \ref{sec:DefiningInHZ} the very definition of
the InHZ depends on general properties of life as a complex physical
phenomenon despite its specific traits.  However, it is also clear
that different ``models of life'' could be characterized by different
optimum physical conditions where it could thrive (see for instance
the differences between the InHZ limits of the DW models in Figure
\ref{fig:DWInHZ}).  It would be thus obvious that each model of life
will determine different InHZ limits. However, an analogous situation
arises in the definition of the RHZ when considering for instance
different types of clouds or different geological and geodynamical
conditions.  In analogy to what is done at defining the more general
RHZ, we should choose the most general or common traits we could
expect for most forms of life thriving in the Universe.
%FINAL

But, how many different forms of life could exist out there? Are they
knowable even in principle? Although we are far from solving these
questions, we can still make some efforts for calculating the InHZ
limits of the Solar System and elsewhere, at least for the type of
life we know on Earth.
%FINAL

Even if we calculate an ``Earth-life-like'' InHZ in an extrasolar
planetary system, it would be possible that other models of life (for
instance ``extremophiles biotas'') able to contribute at establishing
habitable equilibrium states beyond those limits, still exist.  This
case, however, will not be too different from the case when we can
find extremophile organisms that are able to thrive beyond the limits
of the AHZ.  However, as opposed to extremophile organisms, extreme
biotas will thrive in planets actually habitable for a large range of
organisms (extremophiles or not).
%FINAL

One of the most interesting astronomical consequences of introducing
the concept of an InHZ is that habitability instead of being a
prerrequisite for life could be actually a proxy for an inhabited
environment.  In other words, the detection of a habitable planet
could be the confirmation that life actually exist on its surface. The
key property of habitable environments enabling this possibility is
its intrinsic instability.  Let us illustrate this with an important
example: the existence of a hydrosphere.
%FINAL

Since the seminal works on planetary habitability by \citet{shapley53}
and \citet{Hart79}, the presence of a liquid hydrosphere has been
regarded as a prerequisite for life as we know it.  However it has
been shown that a planetary equilibrium state including a liquid
hydrosphere is highly unstable \citep{Gorshkov02, Gorshkov04}.  In
such a state and in the absence of powerful regulating feedbacks, the
system will rapidly make a transition to equilibrium stable states
where the hydrosphere evaporates causing a catastrophic greenhouse
effect or it completely freezes out.  Obviosuly both states are
prohibitive for life. The origin of such instability is the positive
feedback affecting the equilibrium amount of atmospheric water vapour
and its greenhouse effect.  In the presence of a hydrosphere,
evaporation will increase the atmopsheric water vapour thus increasing
the greenhouse effect and surface temperatures that at its turn
increases even more the evaporation rate and so on.  On the other hand
frozen water have a larger albedo than liquid water which tends to
reduce average temperatures increasing the frozen water in the planet
and so on.
%FINAL

\cite{Gorshkov04} calculated that if left unregulated, the Earth's
hydrosphere will be fully evaporated or frozen in less than $10^4$
years.  Therefore, in order to maintain a global hydrosphere during
geological timescales, the Earth could have required powerful
regulatory mechanisms.  We have already argued how life overcomes by
many orders of magnitude the reglatory power of other abiotic
processes.  In this sense, and as stated before, the presence of a
liquid hydrosphere instead of being a prerrequisite for life can be
actually regarded as a proxy for the presence of life itself.  In
practical terms, detecting oceans and other large masses of liquid
water in the surface of extrasolar planets, instead of pointing to the
possibility that life could thrive on the planet, would be actually a
signature of the actual presence of life.  Oceans could be the
ultimate biosignature.
%FINAL

Is the InHZ concept tightly bound to Gaia or BR theories?  Not at all.
Although both theories provide very important arguments and evidences
supporting the definition of an InHZ, stating that life is an
important factor that could not be disregarded at calculating the
equilibrium state of a habitable planet is very different than
assuming that life is the most important one.  Our point here is that
in a complex system such as an inhabited planet taking away any major
component of the system gives you back another planet.  In simple
words, the whole without its 10\% is not 90\% of the whole (here
10\%-90\% could be replaced by 50\%-50\% or 90\%-10\%)
%FINAL

Are the DW model the only way to approach to the quantitative
determination of the InHZ limits? Definitevly no. Actually the best
way to approach to this complex problem is by modelling in full detail
the complex interactions and feedbacks between life and its abiotic
environment on our own Earth as a first example.  Very complex models
of this sort have been developed in the past.  Our task now will be to
apply those models not for studying the Earth but for looking for
other Earths in the Universe.
%FINAL

%%%%%%%%%%%%%%%%%%%%%%%%%%%%%%%%%%%%%%%%%%%%%%%%%%%%%%%%%%%%%%%%%%%%%%%%%
\section{Summary and Conclusions}
\label{sec:Conclusions}
%%%%%%%%%%%%%%%%%%%%%%%%%%%%%%%%%%%%%%%%%%%%%%%%%%%%%%%%%%%%%%%%%%%%%%%%%

We have presented here theoretical arguments supporting the idea that
life cannot be excluded when finding the plausible equilibrium states
that define the limits of the Habitable Zone.  Since our final goal at
searching for habitable planets is precisely looking for the inhabited
ones, turning from the traditional definition of an AHZ to a more
general InHZ is mandatory.  The arguments presented here were based on
mounting observational evidence as well as on theories developed along
the last decades, supporting the idea that life have a non-negligible
effect on the environment of the only habitable planet we know so far:
the Earth.  Although including life in all its complexity in a
realistic model of the environment of any inhabited planet is
challenging, we have shown trhough simple albeit illustrative
conceptual and numerical experiments that it can be achieved.  More
importantly, we showed that life is able to substantially modify the
limits of the otherwise uninhabited AHZ and therefore, potentially
expand the region in the parameter space where we are presently
searching for it. Together, the theoretical arguments, the
observational evidences, and the simple examples provided here,
constitute a general conceptual framework on which more complex models
of the InHZ can be developed.
%FINAL

The InHZ, as defined here, is the region where the interaction between
life and its abiotic planetary environment supports the necessary
physical conditions for the very existence and persistence of life
itself.  This concept is in starking contrast with the definition of
an AHZ which is commonly used in astrobiology and exoplanetary
research.  Our work emphasizes the fact that habitability is an
emergent property of the complex biota-environment coupled system and
not simply a physical prerrequisite entirely determined by
astrophysical, geophysical and other abiotic factors.
%FINAL

We did not address here the problem of the origin of life at defining
the InHZ in the same way as the ideas of an abiotic habitability have
never addressed the problem of the origin of liquid water.  In other
words, we argue that answering the question ``which comes first, life
or habitability?'' is analogous to trying to answer the question
``which comes first, liquid water or habitable surface
temperatures?''.  Complex systems such as habitable planets or in
general planetary environments are characterized by this sort of what
should be considered fake egg-and-chicken paradoxes.  The origin of
emergent properties in complex system does not require simple
sequential explanations.
%FINAL

Our definition of an InHZ should not be confused with a definition
based on the capacity of extremophiles organisms to thrive under
conditions beyond the limits of the living organisms on Earth.  By
definition, an inhabited habitable planet should guarantee habitable
conditions to all organisms able to thrive in the range of
environmental conditions characterizing the equilibrium state of the
planet.  If these equilibrium conditions are extreme for Earth
organisms it does not make life in this planet extremophilic but just
different.  In the same line of reasoning, our InHZ definition does
not consider the case of hidden biospheres (e.g. life thriving in the
solid or liquid planetary interior). As usual at defining the InHZ we
are looking for life able to produce detectable signatures in the
planetary atmosphere and/or its surface.
%FINAL

Our aim here was not to provide limits of the InHZ but to pose the
question {\it what if} habitable zone models are excluding a key
component of the planetary environment. In that sense our central
point should be read not as a prove but as a question.  Paraphrasing
the Einstein's quote at the beginning of this paper ``we should make
things as simple as possible, but not simpler''.  When dealing with
habitability we should consider as few factors as possible but not
fewer. Life is certainly an unavoidable factor.
%FINAL

The InHZ concept could be further explored in several directions and
may serve as a conceptual framework for developing more realistic
models of planetary habitability. For instance, we can try to
introduce already known biotic feedbacks in some widely accepted
models of abiotic habitability.  On the other hand we can improve the
most simple DW models including a more realistic treatment of the
response of the atmosphere to solar forcing.  Although the models by
\citet{Salazar09} were aimed in that direction, further efforts to
improve their atmospheric model should be pursued.  Evolution is a key
process for life and its role at determining the way as biota alter
its environment would also be a key step towards more realistic InHZ
models.
%FINAL

%%%%%%%%%%%%%%%%%%%%%%%%%%%%%%%%%%%%%%%%%%%%%%%%%%%%%%%%%%%%%%%%%%%%%%
%Acknowledgements
%%%%%%%%%%%%%%%%%%%%%%%%%%%%%%%%%%%%%%%%%%%%%%%%%%%%%%%%%%%%%%%%%%%%%%
\begin{acknowledgements}
% TEXT
% \section*{Acknowledgments}

We want to especially thank to our colleagues Victor Gorshkov,
Anastassia Makarieva, Rene Heller, Dave Waltham and Peter Bunyard for
a preliminary revision of the manuscript.  Their insightful comments,
corrections and clever questions were fundamental at determining the
final version of the manuscript.  JZ thanks the AMEBA group for
inspiring some of the key ideas in this work and for allowing us to
discuss some of them in its weekly meetings.  This work has been done
with the financial support of CODI-UdeA.
\end{acknowledgements}

%%%%%%%%%%%%%%%%%%%%%%%%%%%%%%%%%%%%%%%%%%%%%%%%%%%%%%%%%%%%%%%%%%%%%%
%Disclosure statement
%%%%%%%%%%%%%%%%%%%%%%%%%%%%%%%%%%%%%%%%%%%%%%%%%%%%%%%%%%%%%%%%%%%%%%
\section*{Auhtor disclosure statement}

All the authors manifest that no competing financial interests exist
in connection with the ideas published in this manuscript.

%%%%%%%%%%%%%%%%%%%%%%%%%%%%%%%%%%%%%%%%%%%%%%%%%%%%%%%%%%%%%%%%%%%%%%%%%%%%
%BIBLIOGRAPHY
%%%%%%%%%%%%%%%%%%%%%%%%%%%%%%%%%%%%%%%%%%%%%%%%%%%%%%%%%%%%%%%%%%%%%%%%%%%%
\bibliography{bibliography}

\begin{thebibliography}{85}
\providecommand{\natexlab}[1]{#1}
\providecommand{\url}[1]{{\tt #1}}
\providecommand{\urlprefix}{URL }
\expandafter\ifx\csname urlstyle\endcsname\relax
  \providecommand{\doi}[1]{doi:\discretionary{}{}{}#1}\else
  \providecommand{\doi}{doi:\discretionary{}{}{}\begingroup
  \urlstyle{rm}\Url}\fi

\bibitem[{Andreae et~al.(2004)Andreae, Rosenfeld, Artaxo, Costa, Frank, Longo,
  and Silva-Dias}]{Andreae04}
Andreae, M., Rosenfeld, D., Artaxo, P., Costa, A., Frank, G., Longo, K., and
  Silva-Dias, M.: Smoking rain clouds over the Amazon, Science, 303,
  1337--1342, 2004.

\bibitem[{Arneth et~al.(2010)Arneth, Harrison, Zaehle, Tsigaridis, Menon,
  Bartlein, Feichter, Korhola, Kulmala, O'donnell et~al.}]{Arneth10}
Arneth, A., Harrison, S., Zaehle, S., Tsigaridis, K., Menon, S., Bartlein, P.,
  Feichter, J., Korhola, A., Kulmala, M., O'donnell, D., et~al.: Terrestrial
  biogeochemical feedbacks in the climate system, Nature Geoscience, 3,
  525--532, 2010.

\bibitem[{Batalha et~al.(2013)Batalha, Rowe, Bryson, Barclay, Burke, Caldwell,
  Christiansen, Mullally, Thompson, Brown et~al.}]{Batalha13}
Batalha, N.~M., Rowe, J.~F., Bryson, S.~T., Barclay, T., Burke, C.~J.,
  Caldwell, D.~A., Christiansen, J.~L., Mullally, F., Thompson, S.~E., Brown,
  T.~M., et~al.: Planetary Candidates Observed by Kepler. III. Analysis of the
  First 16 Months of Data, The Astrophysical Journal Supplement Series, 204,
  24, 2013.

\bibitem[{Beerling(2005)}]{Beerling05}
Beerling, D.: Leaf evolution: gases, genes and geochemistry, Annals of Botany,
  96, 345--352, 2005.

\bibitem[{Bonan(2008)}]{Bonan08}
Bonan, G.: Forests and climate change: forcings, feedbacks, and the climate
  benefits of forests, science, 320, 1444--1449, 2008.

\bibitem[{{Boyle} et~al.(2011){Boyle}, {Lenton}, and {Watson}}]{Boyle11}
{Boyle}, R., {Lenton}, T., and {Watson}, A.: {Symbiotic
  physiologypromoteshomeostasisinDaisyworld}, Journal ofTheoreticalBiology,
  274, 170--182, 2011.

\bibitem[{Caldeira et~al.(1992)Caldeira, Kasting et~al.}]{CaldeiraKasting92}
Caldeira, K., Kasting, J., et~al.: The life span of the biosphere revisited,
  Nature, 360, 721--723, 1992.

\bibitem[{{Carter} and {Prince}(1981)}]{Carter81}
{Carter}, R.~N. and {Prince}, S.~D.: {Epidemic models used to explain
  biogeographical distribution limits}, \nat, 293, 644--645, 1981.

\bibitem[{Cohen and Rich(2000)}]{Cohen00}
Cohen, J. and Rich, A.: Interspecific competition affects temperature stability
  in Daisyworld, Tellus B, 52, 980--984, 2000.

\bibitem[{DeLeon-Rodriguez et~al.(2013)DeLeon-Rodriguez, Lathem, Rodriguez-R,
  Barazesh, Anderson, Beyersdorf, Ziemba, Bergin, Nenes, and
  Konstantinidis}]{Deleon13}
DeLeon-Rodriguez, N., Lathem, T.~L., Rodriguez-R, L.~M., Barazesh, J.~M.,
  Anderson, B.~E., Beyersdorf, A.~J., Ziemba, L.~D., Bergin, M., Nenes, A., and
  Konstantinidis, K.~T.: Microbiome of the upper troposphere: Species
  composition and prevalence, effects of tropical storms, and atmospheric
  implications, Proceedings of the National Academy of Sciences, 110,
  2575--2580, 2013.

\bibitem[{{Dole}(1964)}]{Dole64}
{Dole}, S.~H.: {Habitable planets for man}, Cambridge, 1964.

\bibitem[{Dyke et~al.(2011)Dyke, Gans, and Kleidon}]{Dyke11}
Dyke, J., Gans, F., and Kleidon, A.: Towards understanding how surface life can
  affect interior geological processes: a non-equilibrium thermodynamics
  approach, Earth System Dynamics, 2, 139--160, 2011.

\bibitem[{Foley et~al.(2003)Foley, Costa, Delire, Ramankutty, and
  Snyder}]{Foley03}
Foley, J.~A., Costa, M.~H., Delire, C., Ramankutty, N., and Snyder, P.: Green
  surprise? How terrestrial ecosystems could affect earth's climate, Frontiers
  in Ecology and the Environment, 1, 38--44, 2003.

\bibitem[{{Franck}(2001)}]{Franck01b}
{Franck}, S.: {Global water cycle and Earth's thermal evolution}, Journal of
  Geodynamics, 32, 231--246, 2001.

\bibitem[{{Franck} et~al.(1999){Franck}, {Kossacki}, and {Bounama}}]{Franck99}
{Franck}, S., {Kossacki}, K., and {Bounama}, C.: {Modelling the global carbon
  cycle for the past and future evolution of the earth system}, Chemical
  Geology, 159, 305--317, 1999.

\bibitem[{{Franck} et~al.(2000{\natexlab{a}}){Franck}, {Block}, {von Bloh},
  {Bounama}, {Schellnhuber}, and {Svirezhev}}]{Franck00b}
{Franck}, S., {Block}, A., {von Bloh}, W., {Bounama}, C., {Schellnhuber},
  H.-J., and {Svirezhev}, Y.: {Habitable zone for Earth-like planets in the
  solar system}, \planss, 48, 1099--1105, 2000{\natexlab{a}}.

\bibitem[{{Franck} et~al.(2000{\natexlab{b}}){Franck}, {von Bloh}, {Bounama},
  {Steffen}, {Sch{\"o}nberner}, and {Schellnhuber}}]{Franck00a}
{Franck}, S., {von Bloh}, W., {Bounama}, C., {Steffen}, M., {Sch{\"o}nberner},
  D., and {Schellnhuber}, H.-J.: {Determination of habitable zones in
  extrasolar planetary systems: Where are Gaia's sisters?}, \jgr, 105,
  1651--1658, 2000{\natexlab{b}}.

\bibitem[{{Franck} et~al.(2001){Franck}, {Block}, {Bloh}, {Bounama}, {Garrido},
  and {Schellnhuber}}]{Franck01a}
{Franck}, S., {Block}, A., {Bloh}, W., {Bounama}, C., {Garrido}, I., and
  {Schellnhuber}, H.-J.: {Planetary habitability: is Earth commonplace in the
  Milky Way?}, Naturwissenschaften, 88, 416--426, 2001.

\bibitem[{Gorshkov(1995)}]{Gorshkov95}
Gorshkov, V.: Physical and biological bases of life stability. Man, biota,
  environment, Springer, 1995.

\bibitem[{Gorshkov and Makarieva(2001)}]{Gorshkov01}
Gorshkov, V. and Makarieva, A.: On the possibility of physical
  self-organization of biological and ecological systems, Doklady Biological
  Sciences, 378, 258--261, 2001.

\bibitem[{Gorshkov and Makarieva(2002)}]{Gorshkov02}
Gorshkov, V. and Makarieva, A.: Greenhouse effect dependence on atmospheric
  concentrations of greenhouse substances and the nature of climate stability
  on Earth, Atmospheric Chemistry and Physics Discussions, 2, 289--337, 2002.

\bibitem[{Gorshkov et~al.(2000)Gorshkov, Gorshkov, and Makarieka}]{Gorshkov00}
Gorshkov, V.~G., Gorshkov, V.~V., and Makarieka, A.~M.: {Biotic regulation of
  the environment}, Springer, 2000.

\bibitem[{Gorshkov et~al.(2004)Gorshkov, Makarieva, and Gorshkov}]{Gorshkov04}
Gorshkov, V.~G., Makarieva, A.~M., and Gorshkov, V.~V.: {Revising the
  fundamentals of ecological knowledge: {T}he biota-environment interaction},
  Ecological Complexity, 1, 17--36, 2004.

\bibitem[{{Hart}(1979)}]{Hart79}
{Hart}, M.~H.: {Habitable Zones about Main Sequence Stars}, \icarus, 37,
  351--357, 1979.

\bibitem[{{Harting} and {Lovelock}(1996)}]{Harding96}
{Harting}, S. and {Lovelock}, J.: {Exploiter-mediated coexistence and
  frequency-dependent selection in a numerical model of biodiversity}, J.
  Theoretical Biology, 182, 109--116, 1996.

\bibitem[{{Hitchcock} and {Lovelock}(1967)}]{Hitchcock67}
{Hitchcock}, D.~R. and {Lovelock}, J.~E.: {Life detection by atmospheric
  analysis}, \icarus, 7, 149--159, 1967.

\bibitem[{H{\"o}ning et~al.(2013)H{\"o}ning, Hansen-Goos, Airo, and
  Spohn}]{Honing13}
H{\"o}ning, D., Hansen-Goos, H., Airo, A., and Spohn, T.: Biotic vs. abiotic
  Earth: A model for mantle hydration and continental coverage, Planetary and
  Space Science, 2013.

\bibitem[{Houghton et~al.(2001)Houghton, Ding, Griggs, Noguer, van~der LINDEN,
  Dai, Maskell, and Johnson}]{Houghton01}
Houghton, J., Ding, Y., Griggs, D., Noguer, M., van~der LINDEN, P., Dai, X.,
  Maskell, K., and Johnson, C.: Climate change 2001: the scientific basis,
  Cambridge University Press Cambridge, 2001.

\bibitem[{Hutjes et~al.(1998)Hutjes, Kabat, Running, Shuttleworth, Field, Bass,
  da~Silva~Dias, Avissar, Becker, Claussen et~al.}]{Hutjes98}
Hutjes, R., Kabat, P., Running, S., Shuttleworth, W., Field, C., Bass, B.,
  da~Silva~Dias, M., Avissar, R., Becker, A., Claussen, M., et~al.: Biospheric
  aspects of the hydrological cycle, Journal of Hydrology, 212, 1--21, 1998.

\bibitem[{Jasechko et~al.(2013)Jasechko, Sharp, Gibson, Birks, Yi, and
  Fawcett}]{Jasechko13}
Jasechko, S., Sharp, Z.~D., Gibson, J.~J., Birks, S.~J., Yi, Y., and Fawcett,
  P.~J.: Terrestrial water fluxes dominated by transpiration, Nature, 2013.

\bibitem[{{Kasting}(2010)}]{Kasting10}
{Kasting}, J.: {How to Find a Habitable Planet}, Princeton University Press,
  2010.

\bibitem[{{Kasting} et~al.(1993){Kasting}, {Whitmire}, and
  {Reynolds}}]{Kasting93}
{Kasting}, J.~F., {Whitmire}, D.~P., and {Reynolds}, R.~T.: {Habitable Zones
  around Main Sequence Stars}, \icarus, 101, 108--128, 1993.

\bibitem[{Keeling et~al.(1996)Keeling, Chin, and Whorf}]{Keeling96}
Keeling, C., Chin, J., and Whorf, T.: Increased activity of northern vegetation
  inferred from atmospheric CO2 measurements, Nature, 382, 146--149, 1996.

\bibitem[{Keeling(2008)}]{Keeling08}
Keeling, R.: Atmospheric science. Recording Earth's vital signs., Science, 319,
  1771, 2008.

\bibitem[{Kitzmann et~al.(2010)Kitzmann, Patzer, von Paris, Godolt, Stracke,
  Gebauer, Grenfell, and Rauer}]{Kitzmann10}
Kitzmann, D., Patzer, A., von Paris, P., Godolt, M., Stracke, B., Gebauer, S.,
  Grenfell, J., and Rauer, H.: Clouds in the atmospheres of extrasolar planets.
  I. Climatic effects of multi-layered clouds for Earth-like planets and
  implications for habitable zones, arXiv preprint arXiv:1002.2927, 2010.

\bibitem[{Kleidon(2009)}]{Kleidon09a}
Kleidon, A.: Maximum entropy production and general trends in biospheric
  evolution, Paleontological Journal, 43, 980--985, 2009.

\bibitem[{Kleidon(2010{\natexlab{a}})}]{Kleidon10}
Kleidon, A.: Life, hierarchy, and the thermodynamic machinery of planet Earth,
  Physics of life reviews, 7, 424--460, 2010{\natexlab{a}}.

\bibitem[{Kleidon(2010{\natexlab{b}})}]{Kleidon10b}
Kleidon, A.: Life as the major driver of planetary geochemical disequilibrium.
  Reply to comments on Life, hierarchy, and the thermodynamic machinery of
  planet Earth, Physics of life reviews, 7, 473--476, 2010{\natexlab{b}}.

\bibitem[{Kleidon(2012)}]{Kleidon12}
Kleidon, A.: How does the Earth system generate and maintain thermodynamic
  disequilibrium and what does it imply for the future of the planet?,
  Philosophical Transactions of the Royal Society A: Mathematical, Physical and
  Engineering Sciences, 370, 1012--1040, 2012.

\bibitem[{Kleidon and Lorenz(2005)}]{Kleidon05}
Kleidon, A. and Lorenz, R.~D.: Non-equilibrium thermodynamics and the
  production of entropy: life, earth, and beyond, Springer, 2005.

\bibitem[{{Kopparapu} et~al.(2013){Kopparapu}, {Ramirez}, {Kasting}, {Eymet},
  {Robinson}, {Mahadevan}, {Terrien}, {Domagal-Goldman}, {Meadows}, and
  {Deshpande}}]{Kopparapu13}
{Kopparapu}, R.~K., {Ramirez}, R., {Kasting}, J.~F., {Eymet}, V., {Robinson},
  T.~D., {Mahadevan}, S., {Terrien}, R.~C., {Domagal-Goldman}, S., {Meadows},
  V., and {Deshpande}, R.: {Habitable Zones around Main-sequence Stars: New
  Estimates}, \apj, 765, 131, 2013.

\bibitem[{{Lenton}(1998)}]{Lenton98}
{Lenton}, T.: {Gaia and natural selection}, Nature, 934, 439--447, 1998.

\bibitem[{Lenton(2002)}]{Lenton02}
Lenton, T.: {Testing Gaia: the effect of life on Earth's habitability and
  regulation}, Climatic Change, 52, 409--422, 2002.

\bibitem[{Lenton(2004)}]{Lenton04}
Lenton, T.: {Clarifying Gaia: regulation with or without natural selection},
  Scientists Debate Gaia, pp. 15--25, 2004.

\bibitem[{{Lenton} and {Lovelock}(2000)}]{Lenton00}
{Lenton}, T. and {Lovelock}, J.: {Daisyworld is Darwinian:Constraints on
  Adaptation are Important for Planetary Self-Regulation}, J. Theoretical
  Biology, 206, 109--114, 2000.

\bibitem[{{Lenton} and {Lovelock}(2001)}]{Lenton01b}
{Lenton}, T. and {Lovelock}, J.: {Daisyworld revisited: Quantifying biological
  effects on planetary self-regulation}, Tellus, 53B, 288--305, 2001.

\bibitem[{Lenton and von Bloh(2001)}]{Lenton01}
Lenton, T. and von Bloh, W.: Biotic feedback extends the life span of the
  biosphere, Geophys. Res. Lett, 28, 1715--1718, 2001.

\bibitem[{Lenton and Wilkinson(2003)}]{Lenton03}
Lenton, T. and Wilkinson, D.: Developing the Gaia Theory. A Response to the
  Criticisms of Kirchner and Volk, Climatic Change, 58, 1--12, 2003.

\bibitem[{Lorenz(2002)}]{Lorenz02}
Lorenz, R.~D.: Planets, life and the production of entropy, International
  Journal of Astrobiology, 1, 3--13, 2002.

\bibitem[{Lovelock(1992)}]{Lovelock92}
Lovelock, J.: A numerical model for biodiversity, Philosophical Transactions of
  the Royal Society of London. Series B: Biological Sciences, 338, 383--391,
  1992.

\bibitem[{Lovelock(1995)}]{Lovelock95}
Lovelock, J.: The ages of Gaia: a biography of our living earth, Oxford
  University Press, 1995.

\bibitem[{Lovelock(1979)}]{Lovelock79}
Lovelock, J.~E.: {Gaia: A New Look at Life on Earth}, Oxford University Press,
  1979.

\bibitem[{{Lovelock}(1989)}]{Lovelock89}
{Lovelock}, J.~E.: {Geophysiology}, Transactions of the Royal Society of
  Edinburgh: Earth Sciences, 80, 169--175, 1989.

\bibitem[{Lovelock(2009)}]{Lovelock09}
Lovelock, J.~E.: {The vanishing face of gaia: a final warning}, Basic Books,
  New York, 2009.

\bibitem[{{Lovelock} and {Margulis}(1974)}]{Lovelock74}
{Lovelock}, J.~E. and {Margulis}, L.: {Homeostatic tendencies of the Earth's
  atmosphere}, Origins of Life, 5, 93--103, 1974.

\bibitem[{Lyons(2002)}]{Lyons02}
Lyons, T.: Clouds prefer native vegetation, Meteorology and Atmospheric
  Physics, 80, 131--140, 2002.

\bibitem[{Makarieva et~al.(2006)Makarieva, Gorshkov, Li, and
  Chown}]{Makarieva06}
Makarieva, A., Gorshkov, V., Li, B., and Chown, S.: Size-and
  temperature-independence of minimum life-supporting metabolic rates,
  Functional Ecology, 20, 83--96, 2006.

\bibitem[{Makarieva et~al.(2007)Makarieva, Gorshkov et~al.}]{Makarieva07}
Makarieva, A., Gorshkov, V., et~al.: Biotic pump of atmospheric moisture as
  driver of the hydrological cycle on land, Hydrology and Earth System Sciences
  Discussions, 11, 1013--1033, 2007.

\bibitem[{Makarieva et~al.(2010)Makarieva, Gorshkov, Sheil, Nobre, and
  Li}]{Makarieva10}
Makarieva, A., Gorshkov, V., Sheil, D., Nobre, A., and Li, B.: Where do winds
  come from? A new theory on how water vapor condensation influences
  atmospheric pressure and dynamics, Atmospheric Chemistry and Physics
  Discussions, 10, 2010.

\bibitem[{{Margulis} and {Lovelock}(1974)}]{Margulis74}
{Margulis}, L. and {Lovelock}, J.~E.: {Biological Modulation of the Earth's
  Atmosphere}, \icarus, 21, 471, 1974.

\bibitem[{Meskhidze and Nenes(2006)}]{Meskhidze06}
Meskhidze, N. and Nenes, A.: Phytoplankton and cloudiness in the Southern
  Ocean, Science, 314, 1419--1423, 2006.

\bibitem[{Mischna et~al.(2000)Mischna, Kasting, Pavlov, and
  Freedman}]{Mischna00}
Mischna, M.~A., Kasting, J.~F., Pavlov, A., and Freedman, R.: Influence of
  carbon dioxide clouds on early Martian climate, Icarus, 145, 546--554, 2000.

\bibitem[{Nevison et~al.(1999)Nevison, Gupta, and Klinger}]{Nevison99}
Nevison, C., Gupta, V., and Klinger, L.: Self-sustained temperature
  oscillations on {D}aisyworld, Tellus, 51B, 806--814, 1999.

\bibitem[{Nicolis and Portnow(1973)}]{Nicolis73}
Nicolis, G. and Portnow, J.: Chemical oscillations, Chemical Reviews, 73,
  365--384, 1973.

\bibitem[{Ozawa et~al.(2003)Ozawa, Ohmura, Lorenz, and Pujol}]{Ozawa03}
Ozawa, H., Ohmura, A., Lorenz, R.~D., and Pujol, T.: The second law of
  thermodynamics and the global climate system: a review of the maximum entropy
  production principle, Reviews of Geophysics, 41, 1018, 2003.

\bibitem[{P{\"o}schl et~al.(2010)P{\"o}schl, Martin, Sinha, Chen, Gunthe,
  Huffman, Borrmann, Farmer, Garland, Helas et~al.}]{Poschl10}
P{\"o}schl, U., Martin, S., Sinha, B., Chen, Q., Gunthe, S., Huffman, J.,
  Borrmann, S., Farmer, D., Garland, R., Helas, G., et~al.: Rainforest aerosols
  as biogenic nuclei of clouds and precipitation in the Amazon, Science, 329,
  1513--1516, 2010.

\bibitem[{Poveda et~al.(2014)Poveda, Jaramillo, and Vallejo}]{Poveda14}
Poveda, G., Jaramillo, L., and Vallejo, L.~F.: Seasonal precipitation patterns
  along pathways of South American low-level jets and aerial rivers, Water
  Resources Research, 2014.

\bibitem[{Ramanathan et~al.(1989)Ramanathan, Cess, Harrison, Minnis, Barkstrom,
  Ahmad, Hartmann et~al.}]{Ramanathan89}
Ramanathan, V., Cess, R., Harrison, E., Minnis, P., Barkstrom, B., Ahmad, E.,
  Hartmann, D., et~al.: Cloud-radiative forcing and climate: Results from the
  Earth Radiation Budget Experiment., Science, 243, 57, 1989.

\bibitem[{Rial et~al.(2004)Rial, Pielke, Beniston, Claussen, Canadell, Cox,
  Held, De~Noblet-Ducoudr{\'e}, Prinn, Reynolds et~al.}]{Rial04}
Rial, J., Pielke, R., Beniston, M., Claussen, M., Canadell, J., Cox, P., Held,
  H., De~Noblet-Ducoudr{\'e}, N., Prinn, R., Reynolds, J., et~al.:
  Nonlinearities, feedbacks and critical thresholds within the Earth's climate
  system, Climatic Change, 65, 11--38, 2004.

\bibitem[{Ridgwell and Zeebe(2005)}]{Ridgwell05}
Ridgwell, A. and Zeebe, R.: The role of the global carbonate cycle in the
  regulation and evolution of the Earth system, Earth and Planetary Science
  Letters, 234, 299--315, 2005.

\bibitem[{Roberston and Robinson(1998)}]{Robertson98}
Roberston, D. and Robinson, J.: Darwinian daisyworld, Journal of Theoretical
  Biology, 195, 129--134, 1998.

\bibitem[{{Rosing} et~al.(2010){Rosing}, {Bird}, {Sleep}, and
  {Bjerrum}}]{Rosing10}
{Rosing}, M.~T., {Bird}, D.~K., {Sleep}, N.~H., and {Bjerrum}, C.~J.: {No
  climate paradox under the faint early Sun}, \nat, 464, 744--747, 2010.

\bibitem[{Runyan et~al.(2012)Runyan, D'Odorico, and Lawrence}]{Runyan12}
Runyan, C., D'Odorico, P., and Lawrence, D.: Physical and biological feedbacks
  of deforestation, Reviews of Geophysics, 50, RG4006, 2012.

\bibitem[{{Sagan} and {Mullen}(1972)}]{Sagan72}
{Sagan}, C. and {Mullen}, G.: {Earth and Mars: Evolution of Atmospheres and
  Surface Temperatures}, Science, 177, 52--56, 1972.

\bibitem[{Salazar and Poveda(2009)}]{Salazar09}
Salazar, J.~F. and Poveda, G.: Role of a simplified hydrological cycle and
  clouds in regulating the climate-biota system of {D}aisyworld, Tellus B, 61,
  483--497, 2009.

\bibitem[{Schellnhuber(1999)}]{Schellnhuber99}
Schellnhuber, H.: ‘Earth system’analysis and the second Copernican
  revolution, Nature, 402, C19--C23, 1999.

\bibitem[{Schimel(1995)}]{Schimel95}
Schimel, D.: Terrestrial ecosystems and the carbon cycle, Global change
  biology, 1, 77--91, 1995.

\bibitem[{Schr{\"o}dinger(1992)}]{Schrodinger92}
Schr{\"o}dinger, E.: What is life?: With mind and matter and autobiographical
  sketches, Cambridge University Press, 1992.

\bibitem[{{Selsis} et~al.(2007){Selsis}, {Kasting}, {Levrard}, {Paillet},
  {Ribas}, and {Delfosse}}]{Selsis07}
{Selsis}, F., {Kasting}, J.~F., {Levrard}, B., {Paillet}, J., {Ribas}, I., and
  {Delfosse}, X.: {Habitable planets around the star Gliese 581?}, \aap, 476,
  1373--1387, 2007.

\bibitem[{Shapley(1953)}]{shapley53}
Shapley, H.: Climatic change: evidence, causes, and effects, Climatic Change:
  Evidence, Causes, and Effects, 1, 1953.

\bibitem[{Steffen(2004)}]{Steffen04}
Steffen, W.: Global change and the earth system: a planet under pressure,
  Global change-the IGBP series (, 2004.

\bibitem[{{Walker} et~al.(1981){Walker}, {Hays}, and {Kasting}}]{Walker81}
{Walker}, J.~C.~G., {Hays}, P.~B., and {Kasting}, J.~F.: {A negative feedback
  mechanism for the long-term stabilization of the earth's surface
  temperature}, \jgr, 86, 9776--9782, 1981.

\bibitem[{Watson and Lovelock(1983)}]{Watson83}
Watson, A. and Lovelock, J.: Biological homeostasis of the global environment:
  {T}he parable of {D}aisyworld, Tellus, 35B, 284--289, 1983.

\bibitem[{Wood et~al.(2008)Wood, Ackland, Dyke, Williams, and Lenton}]{Wood08}
Wood, A., Ackland, G., Dyke, J., Williams, H., and Lenton, T.: {{D}aisyworld: A
  review}, Reviews of Geophysics, 46, 1--23, 2008.

\bibitem[{{Zuluaga} et~al.(2013){Zuluaga}, {Bustamante}, {Cuartas}, and
  {Hoyos}}]{Zuluaga13}
{Zuluaga}, J.~I., {Bustamante}, S., {Cuartas}, P.~A., and {Hoyos}, J.~H.: {The
  Influence of Thermal Evolution in the Magnetic Protection of Terrestrial
  Planets}, \apj, 770, 23, 2013.

\end{thebibliography}
\bibliographystyle{copernicus}

\addtocounter{figure}{-1}\renewcommand{\thefigure}{\arabic{figure}a}

\end{document}